# Prospective Implications of EUV Coronal Plumes for Magnetic-network Genesis of Coronal Heating, Coronal-hole Solar Wind, and Solar-wind Magnetic-field Switchbacks

Ronald L. Moore[1,2], Sanjiv K. Tiwari[3,4], Navdeep K. Panesar[3,4], & Alphonse C. Sterling[2]

[1] Center for Space Plasma and Aeronomic Research (CSPAR), UAH, Huntsville AL, 35805 USA; ronald.l.moore@nasa.gov
[2] NASA Marshall Space Flight Center, Huntsville, AL 35812, USA
[3] Lockheed Martin Solar Astrophysics Laboratory, 3251 Hanover Street Building 252, Palo Alto, CA 94304, USA
[4] Bay Area Environmental Research Institute, NASA Research Park, Moffett Field, CA 94035, USA; tiwari@baeri.org

**Abstract**

We propose that coronal heating in EUV coronal plumes is weaker, not stronger, than in adjacent non-plume coronal magnetic funnels.  This expectation stems from (i) the observation that an EUV plume is born as the magnetic flux at the foot of the plume's magnetic funnel becomes tightly packed together, and (ii) the observation that coronal heating in quiet regions increases in proportion to the coast-line length of the underlying magnetic network.  We do not rule out the possibility that coronal heating in EUV plumes might be stronger, not weaker, but we point out how the opposite is plausible.  We reason that increasing coronal heating during plume birth would cause co-temporal increasing net upward mass flux in the plume, whereas decreasing coronal heating during plume birth would cause co-temporal net downward mass flux in quiet-region plumes and co-temporal decrease in net upward mass flux or even net downward mass flux in coronal-hole plumes.  We further reason that conclusive evidence of weaker coronal heating in EUV plumes would strengthen the possibility that magnetic twist waves from fine-scale magnetic explosions at the edges of the magnetic network (1) power much of the coronal heating in quiet regions, and (2) power most of the coronal heating and solar wind acceleration in coronal holes, with many twist waves surviving to become magnetic-field switchbacks in the solar wind from coronal holes.

## 1. Introduction and Overview

   This paper is dedicated to Peter Sturrock.  Peter introduced Ron Moore to solar physics and solar physics research.  That led to Ron's PhD Dissertation, "The Structure and Heating of the Chromosphere-Corona Transition Region," the insights of which are laid out in Moore & Fung (1972) and are germane to the considerations presented here.
   It is generally accepted that the inner solar corona – at heights within a tenth or so of a solar radius above the Sun's much denser and much cooler ($T < 10^4$ K) visible surface layers, the photosphere and chromosphere – is a roughly hydrostatic gravitationally bound atmosphere.  It is also generally thought that it is essentially due to the corona's mega-Kelvin temperature that the corona's outer fringes – beyond roughly two solar radii from Sun center – continually escape to become the heliosphere-making solar wind that blows out far beyond the planets, finally merging with the interstellar wind (Parker 1958, 1963; Cranmer & Winebarger 2019). In the quasi-static inner corona, the temperature is (1 – 2) million K in coronal holes and quiet regions and twice or so hotter in active regions (Withbroe & Noyes 1977). That is, everywhere on the Sun, the corona close to the Sun's visible surface layers is at least a hundred times hotter than those layers. By the second law of thermodynamics, to sustain that temperature difference, there must be transmitted outward from the surface layers enough of some form of nonthermal energy that is dissipated into heat in the corona. That much is certain, but learning how the Sun does that remains a long-standing flagship endeavor of solar astrophysics. Solar-stellar coronal heating is understood in the above broad way, but what the specific main governing mechanisms are remains murky and hotly debated (Cranmer & Winebarger 2019).

For fifty years – from Skylab on – from soft X-ray images and coronal EUV images of the whole Earth-facing inner corona together with full-disk magnetograms, it has been obvious that the solar corona is heated and shaped by magnetic field rooted in the photosphere (e.g., Vaiana & Rosner 1978; Falconer et al 1997; Moore et al 1999, 2022). Generally, the stronger the magnetic field's photospheric strength and flux in a solar region the brighter and hotter the soft X-ray emitting plasma in that field's coronal body. [Magnetic loops that arch into the corona but have their opposite feet in opposite-polarity sunspot umbrae are a major exception to the rule. They have no coronal X-ray and coronal EUV emitting plasma even though their foot fields are stronger than in bright coronal loops (Tiwari et al 2017).] Evidently, the flux of nonthermal energy that keeps the corona hot by its dissipation is transmitted into the corona magnetically, and generally increases with the field's photospheric strength and flux. It is now well accepted that most of that nonthermal energy is in waves and/or quasi-static twists somehow presently and/or previously put into the magnetic field by magnetoconvection in and/or below the photosphere (e.g., Parker 1983; Cranmer & Winebarger 2019). Exactly when, where and how the magnetoconvection does that remain open questions (e.g., Tiwari et al 2017; Cranmer & Winebarger 2019; Pontin & Hornig 2020). [That solar-stellar coronal heating is driven by magnetoconvection is nicely confirmed by the observational finding of Tiwari et al (2017) that coronal heating is strongly suppressed in umbra-to-umbra coronal magnetic loops because the field in their feet is so strong that it strongly suppresses magnetoconvection.]

In quiet regions and coronal holes, the supergranule convection cells – each roughly 30,000 km in diameter – keep the photospheric magnetic flux swept into clumps that sit on the downflows along the cell edges, forming the so-called magnetic network that roughly maps the network of supergranule convection downflows (Beckers 1977). A majority of the network magnetic flux in a coronal hole has the same polarity, and the coronal hole is filled with field that is rooted in the majority-polarity flux and has been opened and is kept open by the solar wind (Zirker 1977). Coronal holes on the central disk typically span at least several supergranules. Often, in a central-disk quiet region, throughout an area spanning at least several supergranules, the magnetic network is quasi-unipolar to the degree typical of coronal holes. In such quasi-unipolar quiet regions, the overlying corona is presumably filled – to heights comparable to the span of the quasi-unipolar quiet region – with legs of far-reaching closed magnetic field loops that have one foot in a majority-polarity flux clump in the quasi-unipolar quiet region's magnetic network and their other feet in opposite-polarity flux clumps in remote quiet regions. In such quasi-unipolar quiet regions and in coronal holes, the magnetic field filling the inner corona funnels down into majority-polarity magnetic flux in the magnetic network. In each magnetic funnel there is the funnel's so-called chromosphere-corona transition region that spans the height interval between the $10^4$ K top of the chromosphere in the funnel and the bottom of the corona in the funnel. In this paper, following Moore & Fung (1972) and Mariska (1992), we take the bottom of the corona in each funnel to be at the height in the funnel at which the temperature reaches $10^6$ K.

The magnetic field filling each magnetic funnel in a coronal hole or quasi-unipolar quiet region is rooted in a majority-polarity flux clump or lane of the magnetic network. Some of the majority-polarity flux on the perimeter of the clump or lane can be the majority-polarity feet of short loops to nearby minority-polarity flux in the interiors of the network cells abutting the flux clump or lane (as indicated in Figure 1). The rest of the clump's or lane's magnetic flux is in the foot of the funnel. Because the funnel is the lower leg and foot of a far-reaching magnetic flux tube, the flux in the funnel's foot equals the flux at every height in the funnel and in the extension of that flux tube farther up into the corona. Above the chromosphere, at heights where the plasma pressure is negligible relative to the magnetic pressure in the magnetic funnels and short loops, the magnetic field in each funnel spreads with height until, in the low corona (at heights above the heights of the short loops), it merges against neighboring funnels and is in magnetic pressure balance with them (as in Figure 1). That means all of the magnetic funnels in a coronal hole or quasi-unipolar quiet region have – at any height where they press against each other – the same magnetic pressure ($B^2/8\pi$) and hence the same magnetic field strength (B). Hence, the greater a funnel's magnetic flux (B x A), the greater the area (A) of the funnel's coronal top relative to the areas of the coronal tops of the other magnetic funnels in the coronal hole or quasi-unipolar quiet region.

Often, some of the magnetic funnels that fill the low corona in a coronal hole or quasi-unipolar quiet region on the central disk are seen as noticeably bright EUV coronal plumes in full-disk images of EUV emission from solar plasma at temperatures of the hotter transition region and the low corona, temperatures near $10^6$ K. Relative to the full-disk images from other EUV passbands of the Atmospheric Imaging Assembly (AIA) of the Solar Dynamics Observatory (SDO), the full-disk images from AIA's 171 Å passband show these EUV coronal plumes most clearly (Avallone et al 2018). The 171 Å passband is centered on Fe IX emission from plasmas at temperatures around 6 x $10^5$ K (Lemen et al 2012). Co-aligned magnetograms such as from SDO's Helioseismic and Magnetic Imager (HMI) show that each bright EUV coronal plume stands on a robust majority-polarity flux clump of the magnetic network (e.g., Avallone et al 2018).

In many papers on EUV coronal plumes, it has been proposed or supposed – explicitly or tacitly – that the heating of the corona and transition region in a plume's magnetic funnel is greater than in the non-plume magnetic funnels that surround and hem in the plume funnel (Wang 1994, 1998; Wang et al 1997, 2016; Raouafi & Stenborg 2014; Poletto 2015; Avallone et al 2018). That view is consistent with the observation that coronal heating generally increases with increasing photospheric magnetic flux and field strength: the surrounding non-plume funnels are rooted in network flux lanes and clumps that have less magnetic flux and weaker field strength than the robust clump on which the plume stands (see Section 2).

In the present paper, we raise the possibility that, instead of being greater, the heating in the (T > $10^6$ K) coronal extension of an EUV-plume magnetic funnel is less than in the coronal extensions of surrounding non-plume magnetic funnels. If, as is usually assumed, heating is greater in EUV-plume magnetic-funnel coronal extensions than in the coronal extensions of surrounding non-plume magnetic funnels, we reason that EUV plumes – in coronal holes and in quiet regions – should have increasing net upward mass flux when the plume is turning on (becoming bright in EUV emission from plasmas at hotter-transition-region and coronal temperatures). Conversely, if the coronal heating is less in the coronal extensions of EUV-plume magnetic funnels than in the coronal extensions of surrounding non-plume magnetic funnels, we reason that (i) quiet-region plumes should have net downward mass flux when the plume is turning on, and (ii) coronal-hole plumes should have decreasing net upward mass flux – or perhaps even net downward mass flux – when the plume is turning on. We consider what each of these two alternative expectations would favor concerning where and how in the magnetic network magnetoconvection drives coronal heating in quiet regions and coronal holes.

In our view, observation that, while they are growing in brightness and size (becoming brighter, wider, and taller, as in Figure 2), EUV plumes definitely have increasing net upward mass flux would be a basic indication that coronal heating increases in growing EUV plumes. We think that would favor driving of the coronal heating in quiet regions and coronal holes mainly by magnetoconvection inside the flux lanes and clumps of the magnetic network. On the other hand, observation that growing EUV plumes definitely have net downward net mass flux in quiet regions and decreasing net upward or even net downward mass flux in coronal holes would be a basic indication that coronal heating decreases in growing EUV plumes. In our view, that would favor the idea (Moore et al 1999, 2011; Falconer et al 2003) that coronal heating in quiet regions and coronal holes is mostly driven by magnetoconvection at the edges of the magnetic network's lanes and clumps of magnetic flux. We also point out that whether or not network-edge magnetoconvection drives much of the coronal heating in quiet regions and most of the coronal heating and solar-wind acceleration in coronal holes, the magnetic-field switchbacks observed by Parker Solar Probe in the solar wind might be remnants of magnetic twist waves launched from tiny magnetic explosions that are prepared and triggered by network-edge magnetoconvection in coronal holes.

## 2. Pertinent Observed Aspects of EUV Coronal Plumes

Each EUV coronal plume tracked in AIA 171 Å images and co-aligned HMI magnetograms is seen to be born, live, and die in synchrony with the compact majority-polarity magnetic flux clump at its foot (Figure 2). This co-evolution progresses as follows (Wang et al 2016; Avallone et al 2018). The plume becomes first faintly discernible and then gradually grows to full width and luminosity in AIA 171 Å images as majority-polarity network flux, carried by converging supergranular convection flows, coalesces into a robust tightly packed clump that is noticeably larger in area and flux than most other network flux clumps in the coronal hole or quasi-unipolar quiet region. As long as the plume-foot flux clump remains tightly packed together – typically from several hours to about a day – the 171 Å plume's luminosity, width and upward extent remain roughly constant and maximal. Finally, the plume gradually fades to invisibility in AIA 171 Å emission as photospheric convection flows break apart and spread out the plume-foot flux clump.

Each EUV plume can be seen in AIA coronal EUV images from each of two other passbands, namely the 193 Å passband and the 211 Å passband. In these images the plume does not stand out as starkly and brightly as in the plume's AIA 171 Å images, which are of Fe IX emission from plasmas at temperatures around $6 \times 10^5$ K. The 193 Å passband is centered on Fe XII emission from plasmas at temperatures around $1.6 \times 10^6$ K, and the 211 Å passband is centered on Fe XIV emission from plasmas at temperatures around $2 \times 10^6$ K (Lemen et al 2012). In this paper, when the plasma temperature (AIA passband) being observed is not specified, the term "EUV plume" refers to the plume's plasma observed at temperatures within a factor of two of 1 MK, hotter and cooler.

EUV plumes were first observed in the 1970's in Skylab EUV images of coronal holes, and were long considered to occur only in coronal holes (Poletto 2015). Only relatively recently, from AIA's 171 Å, 193 Å, and 211 Å EUV images registered with HMI magnetograms, has it been noticed that features essentially identical to coronal-hole EUV plumes occur in quasi-unipolar quiet regions as well (Wang et al 2016; Avallone et al 2018). For that reason, the measurement of plume plasma temperature and density from imaged EUV spectra has so far been done only for EUV plumes in coronal holes (Poletto 2015). In this work, we do not distinguish between coronal-hole plumes and those that occur in quasi-unipolar quiet regions, except when specifically considering one or the other. When not specified, the terms "EUV plume" and "plume" refer to EUV plumes in either kind of region.

EUV spectroscopic diagnostics of coronal-hole plumes have found the following for plume plasma temperature and density at heights of ~ 70,000 km (Del Zanna et al 2003, 2008; Wilhelm et al 2011; Poletto 2015). The plume plasma's temperature distribution (the plume's differential emission measure) is strongly dominated by a high sharp peak at $8 \times 10^5$ K, which is $(2 – 4) \times 10^5$ K cooler than the temperature of the ambient non-plume coronal-hole plasma. The electron density is $(1 – 2) \times 10^9$ cm$^{-3}$, about a factor of 3 greater than in the ambient coronal hole. These spectroscopically measured values for the plasma temperature and density in coronal-hole EUV plumes are nicely consistent with EUV plumes being more outstandingly bright in AIA 171 Å images than in AIA 193 Å images and AIA 211 Å images.

During the plume's maximal epoch, in HMI magnetograms, the magnetic field strength in the plume-foot flux clump is tens of Gauss on the clump's perimeter and hundreds of Gauss inside the clump (Avallone et al 2018). During this time, in AIA 171 Å images, a plume's upward extent is of order twice the width of the plume's top, and that width ranges among plumes from about 20,000 km to about 40,000 km (in the AIA 171 Å plume images shown in Avallone et al 2018). The fraction of the area of a coronal hole or quasi-unipolar quiet region filled by plume tops is no more than about 0.2 (estimated by visual inspection of AIA 171 Å images of the four coronal holes and four quasi-unipolar quiet regions of the plumes studied by Avallone et al 2018).

In AIA 171 Å images (which have 0.6" pixels and 12-second cadence) the plume funnel shows much fine-scale structure consisting of many standing faint striations, each lasting tens of minutes or less (Avallone et al 2018). Some of these shoot upward from the plume foot and are called jetlets (Raouafi & Stenborg 2014). Many jetlets are rooted at the edge of the majority-polarity plume-foot flux clump, where, in at least some cases, there is a tiny patch of minority-polarity flux. This suggests that plume jetlets are small versions of larger coronal jets that are seated on a polarity inversion line at which minority-polarity flux is cancelling with majority-polarity network flux (based on the observations of Adams et al 2014; Raouafi & Stenborg 2014; Panesar et al 2016, 2018, 2019; McGlasson et al 2019).

Spectroscopic blueshifts of the Ne VIII 770 Å line emission from $T \approx 6 \times 10^5$ K plasma – blueshifts made by 5-10 km s$^{-1}$ upflows – have been observed in some magnetic funnels in a polar coronal hole (Hassler et al 1999; Poletto 2015). Whether these upflows are in only EUV plumes, or in only non-EUV-plume magnetic funnels (inter-plume funnels) between the EUV plumes in the coronal hole, or happen in both, is still an open question (Poletto 2015). Blueshifts of Fe X, XII, and XIII lines from plasma at $(1 – 3) \times 10^6$ K temperatures in EUV plumes in central-disk coronal holes show upflow speeds that increase with height, from about 10 km s$^{-1}$ at heights of about 15,000 km to about 25 km s$^{-1}$ at heights of about 35,000 km (Fu et al 2014; Poletto 2015). EUV plumes also show upward propagating fluctuations that are probably compression waves and/or Alfvenic magnetic-twist waves (DeForest & Gurman 1998; Moore et al 2015; Poletto 2015; Panesar et al 2018).

### 3. Coronal Heating and Coronal Mass in Quiet Regions and Coronal Holes

*3.1. Quasi-unipolar Quiet Regions*

The magnetic field in any quiet region is apparently an ensemble of closed loops that range widely in length and height. For considering coronal heating and coronal mass in quasi-unipolar quiet regions, for simplicity, we take the magnetic field rooted in a quasi-unipolar quiet region to be entirely in closed loops of only two populations: (1) low short loops, and (2) high-reaching long loops (Figure 1a). Each low short loop connects flux in an edge of a majority-polarity flux clump or lane of the network to a small patch of minority-polarity flux in the interior of the adjoining network cell. For simplicity, we assume that none of these short loops arch above the ($T < 10^4$ K) chromosphere in the network cell interiors, and that each has no transition-region-temperature ($10^4 < T < 10^6$ K) plasma and no coronal-temperature ($T > 10^6$ K) plasma in it. (Actual quiet regions and coronal holes have some short loops that arch somewhat higher than in Figure 1a,b and/or do have in them some plasma at transition-region and coronal temperatures. That we are ignoring and not depicting these hotter short loops in Figure 1 makes no difference to our reasoning in this paper concerning the transition region and corona in each magnetic funnel and its high-reaching coronal extension.) The rest of the magnetic field is in the high-reaching long loops, each of which extends high above the chromosphere in the quasi-unipolar quiet region. One end of each of these loops is a magnetic funnel rooted in majority-polarity flux in the magnetic network in the quasi-unipolar quiet region. The other end is rooted in opposite-polarity flux far outside the quasi-unipolar quiet region. In this simple conceptual setup, all of the quasi-unipolar quiet region's plasma at coronal temperatures is in the high-reaching long loops, and all of its plasma at transition-region temperatures is in the transition regions in these loops' magnetic-funnel ends in the quasi-unipolar quiet region. In this setup, the far end of each high-reaching magnetic loop funnels into one or more network concentrations of flux of the polarity opposite that of the majority polarity in the quasi-unipolar quiet region. As in the loop's magnetic funnel in the quasi-unipolar quiet region, each of the loop's far-end magnetic funnels has in it a transition region between the $10^4$ K top of the chromosphere in the foot of the funnel and the $10^6$ K bottom of the loop's corona.

Following Moore & Fung (1972), we assume (1) the corona in each high-reaching loop is heated by dissipation of nonthermal energy that is transmitted from the loop's feet up along the magnetic field into the loop's corona, (2) there is negligible dissipation of the nonthermal energy as it transits through the loop's foot-funnel transition regions, and (3) the plasma is in hydrostatic equilibrium against gravity throughout the loop, at all heights and temperatures (during steady coronal heating). What we have in mind for the nonthermal energy that is dissipated into heat in the loop's corona is free magnetic energy in fine-scale quasi-static magnetic twists and/or in fine-scale transient magnetic twists (Alfvenic magnetic twist waves). We suppose that the magnetic twists are generated by the magnetoconvection in the feet of the magnetic-funnel ends of the high-reaching magnetic loop. The magnetoconvection's generation of appropriate quasi-static magnetic twists is plausibly via magnetic braiding a la Parker (1983). The magnetoconvection's generation of appropriate transient magnetic twist waves is perhaps via turbulent downflow between granules a la van Ballegooijen & Cranmer (2008) and/or via fine-scale magnetic explosions prepared and triggered at the edges of network flux clumps and lanes a la Moore et al (1999, 2011) and Falconer et al (2003).

Our above simple conceptual model for the corona and transition region in a high-reaching magnetic loop in a quasi-unipolar quiet region is the model of Moore & Fung (1972) for the transition region and corona in quiet regions, except their model has no funnel tapering of the magnetic field in the transition region. Even though including the field tapering in their model would quantitatively change the response of the transition region to an increase or decrease in the amount of coronal heating, we see no reason why including the taper would change the basic qualitative character of the response. That is, we assume that how the corona and the funnel-end transition regions in one of our conceptual-model high-reaching coronal loops would respond to a change in the amount of coronal heating is essentially the same as for the Moore & Fung (1972) model.

Because the model high-reaching loop is closed and static, it loses no coronal mass and heat to the solar wind. The steady total heating of the loop's corona is balanced by steady thermal radiation from the loop's coronal plasma in tandem with steady heat conduction into the transition regions in the loop's magnetic-funnel feet. In the model steady-heating equilibrium loop, as in the Moore & Fung (1972) model, the essential function of each loop-end transition region is to balance the conduction heat flux into it from the loop's corona by the thermal radiation of the transition region's plasma. Among hydrostatic equilibrium states of the model high-reaching loop, the greater the heating of the loop's coronal plasma, the greater the temperature of that plasma, and, consequently, the greater the heat flux conducted out of the loop's corona into the loop-end transition regions. In turn, the greater the heat flux from the loop's corona into the loop-end transition regions, the greater the hydrostatic plasma's density must be throughout the loop's corona and loop-end transition regions. That is required for the model loop with greater coronal heating to be hydrostatic throughout, have its loop-end transition regions absorb (by radiating) all of the heat flux coming into them from the loop's corona, and thereby have the loop's total radiation from its corona and transition regions balance the loop's total coronal heating (Moore & Fung 1972).

If there were a step increase in the heating of the model loop's corona, the loop-corona temperature and hence the conduction heat flux from the loop corona into each loop-end transition region would suddenly increase. Initially, there would not be enough plasma in each loop-end transition region to radiate the increase in incoming heat flux. Essentially, the excess heat flux would flow down through each loop-end transition region's (T = $10^4$ K) bottom to heat and drive "chromospheric-evaporation" upflow through the transition region and into the loop corona. This would continue until enough new heated plasma were added to the loop corona and its transition regions that the loop's new greater coronal heating were again balanced by the total radiation of the loop's coronal and transition-region plasma, and the loop were again in hydrostatic equilibrium.

Conversely, if there were a step decrease in the heating of the model loop's corona, the loop-corona temperature and hence the conduction heat flux from the loop corona into each loop-end transition region would suddenly decrease. For the new lower coronal temperature, there would initially be more plasma mass in the loop corona than could be supported in hydrostatic equilibrium, and more plasma in each loop-end transition region than could be kept from cooling by the reduced heat flux from the loop's corona. Essentially, the reduced coronal heating would result in cooling excess plasma draining out of the loop's corona and transition regions down into the (T < $10^4$ K) chromosphere. This would continue until enough cooled-plasma mass were lost from the loop's corona and transition regions that the loop's new reduced coronal heating were again balanced by the loop's total radiation from its corona and transition regions, and the loop were again in hydrostatic equilibrium.

From the above reasoning, we see that during an ongoing increase in heating of the corona in a quasi-unipolar quiet region's model loop, there should be upward mass flux through the loop's magnetic-funnel transition-region ends into the loop's corona. Conversely, during an ongoing decrease in heating of the model loop's corona, there should be downward mass flux out of the loop's corona down through the loop's transition-region ends.

*3.2. Coronal Holes*

The magnetic field filling nearly all of the bottom of the corona in any coronal hole is apparently an ensemble of open-field magnetic funnels rooted in majority-polarity flux clumps and lanes of the coronal hole's quasi-unipolar magnetic network. For considering coronal heating and coronal mass in coronal holes, mimicking our simple conceptual magnetic field setup for quasi-unipolar quiet regions, again for simplicity, we take the magnetic field rooted in a coronal hole to be entirely in magnetic structures of only two populations: (1) low short loops, and (2) funnel-ended open flux tubes of magnetic field that has been opened and is held open by the solar wind (Figure 1b). For simplicity, each short loop connects flux in the edge of a majority-polarity flux clump or lane of the coronal hole's magnetic network to a small patch of minority-polarity flux in the interior of the adjoining network cell, does not arch above the chromosphere, and has in it no plasma of either transition-region or coronal temperature. As in our simple conceptual model quasi-unipolar quiet region, in our simple conceptual model coronal hole all of the transition-region-temperature plasma is in the transition regions in the funnel ends of the open magnetic field, and all of the coronal-temperature plasma is in the open-field coronal extensions of the transition-region-holding magnetic funnels.

In quiet regions, the coronal plasma in any static coronal loop, because the loop is closed, loses no heat or mass to the solar wind, as in our conceptual model for the transition region and corona in quasi-unipolar quiet regions. In stark contrast, in coronal holes the corona's dominant loss of heat and mass is to the solar wind that flows out of the coronal hole. In coronal holes, the corona's total energy loss rate per unit photospheric surface area is of order $10^6$ erg cm$^{-2}$ s$^{-1}$, only about a tenth of which is by coronal radiation and downward heat conduction; the rest goes to the solar wind (Withbroe & Noyes 1977). Of the non-solar-wind energy loss rate ($\sim 10^5$ erg cm$^{-2}$ s$^{-1}$), only about a tenth is by radiation from the coronal plasma, and the rest is by heat conduction into the transition region (Withbroe & Noyes 1977). The T = $10^6$ K top (bottom) of the transition region (corona) is only of order $10^4$ km above the photosphere (about 1.01 $R_{Sun}$ from Sun center) in quiet regions (Moore & Fung 1972; Cranmer & Winebarger 2019), but is of order ten times higher in coronal holes. In coronal holes, the $10^6$ K level is about $10^5$ km above the photosphere (about 1.15 $R_{Sun}$ from Sun center) (Cranmer 2009). At that height in a coronal hole, the electron or proton number density is about 3 x $10^7$ cm$^{-3}$, and the solar wind outflow speed is about 30 km s$^{-1}$ (Cranmer 2009). These two values give the rate of mass flow through the corona into the solar wind in a coronal hole to be about 2 x $10^{-10}$ gm cm$^{-2}$ s$^{-1}$. The same value was estimated by Withbroe & Noyes (1977) for the solar wind mass loss rate in coronal holes.

In our conceptual model coronal hole, we assume the following. 1. Each of the funnel-ended open-field flux tubes (that in aggregate fill the coronal hole's corona and solar wind) has in it (during steady coronal heating and solar-wind acceleration) a steady upflow that enters the T = $10^4$ K bottom of the flux tube's transition region, continues up through the flux tube's transition region and corona, and becomes the solar wind from that flux tube of the coronal hole. The steady mass flow rate (per unit photospheric area of the flux tube's heliocentric solid angle) of the upflow is about 2 x $10^{-10}$ gm cm$^{-2}$ s$^{-1}$. At 1.15 $R_{Sun}$ in the model flux tube, the plasma temperature is $10^6$ K, the electron density is about 3 x $10^7$ cm$^{-3}$, the outflow speed is about 30 km s$^{-1}$ and the magnetic field strength is about 10 G (which is the area-average strength of the radial component of the magnetic field measured in a polar coronal hole by Tsuneta et al 2008). 2. The coronal heating and solar wind acceleration in the flux tube are from dissipation and pressure of free magnetic energy that is transmitted along the magnetic field from the foot of the funnel to the flux tube's corona and solar wind.  3. There is negligible dissipation of the free magnetic energy as it transits through the flux tube's transition region.  The flux of free magnetic energy entering the $10^6$ K bottom of the corona at 1.15 $R_{Sun}$ is $10^6$ erg cm$^{-2}$ s$^{-1}$.  What we have in mind for the free magnetic energy that is dissipated into heat and accelerates the solar wind in the flux tube's corona is the free magnetic energy of transient fine-scale magnetic twists in fine-scale Alfvenic magnetic-twist waves generated by magnetoconvection at the photospheric foot of the flux tube's magnetic field.  Enough such magnetic twist waves might be generated in the interior of the foot flux clump by intergranular turbulent downflow a la van Ballegooijen & Cranmer (2008).  Another idea is that magnetoconvection at the perimeter of the foot flux clump generates enough such magnetic twist waves by preparing and triggering fine-scale magnetic explosions a la Moore et al (1999, 2011, 2015), Falconer et al (2003), and Sterling et al (2020a,b).

In coronal holes, from 1.15 $R_{Sun}$ to 3 $R_{Sun}$ from Sun center, the solar wind outflow speed increases from about 30 km s$^{-1}$ to about 300 km s$^{-1}$ (Cranmer 2009).  In this interval, because the temperature of the outflowing coronal plasma is about 1 x $10^6$ K (Cranmer 2009), the sound speed is about 200 km s$^{-1}$: from 1.15 $R_{Sun}$ to 3 $R_{Sun}$, the solar wind outflow is accelerated from being very subsonic to being mildly supersonic.  Even so, from 1.15 $R_{Sun}$ to 1.5 $R_{Sun}$, the interval in which the outflow speed increases to only about 100 km s$^{-1}$ (Cranmer 2009), the corona in a coronal hole is nearly in hydrostatic equilibrium against gravity: the run of decreasing plasma density with increasing distance is only slightly less steep than that of an isothermal hydrostatic coronal-plasma atmosphere in which the temperature is 1 x $10^6$ K and the electron or proton number density at 1.15 $R_{Sun}$ is 3 x $10^7$ cm$^{-3}$ (Allen 1973; Moore et al 1991).  That is, the observed decrease in plasma density with distance from 1.15 $R_{Sun}$ to 1.5 $R_{Sun}$ suggests that near the bottom of the corona in coronal holes (within a few tenths of a solar radius above 1.15 $R_{Sun}$) the coronal plasma is supported against gravity mostly by the gradient of the plasma's thermal pressure and only at most secondarily by a gradient of the pressure of the magnetic twist waves that we suppose heat the corona and accelerate the solar wind in coronal holes.

At the ($10^6$ K, 1.15 $R_{Sun}$) bottom of the corona in our model funnel-ended open flux tube with its foot in a majority-polarity flux clump of the magnetic network in a coronal hole, the thermal plasma pressure ($\approx 2n_e kT$, where $n_e$ is the electron density and k is the Boltzmann constant) is $\approx$ 9 x $10^{-3}$ dyn cm$^{-2}$.  The flux $F_A$ of free magnetic energy carried by an Alfven wave is $[(\delta B)^2/8\pi]V_A$ (giving $(\delta B)^2/8\pi = F_A/V_A$), where $\delta B$ is the root mean square (rms) amplitude of the wave's magnetic field perturbation, $(\delta B)^2/8\pi$ is both the wave's magnetic pressure and the wave's magnetic energy density, and $V_A$ is the Alfven speed ($V_A \approx B/(4\pi m_p n_e)^{1/2}$, where B is the strength of the magnetic field along which the wave propagates, and $m_p$ is the proton mass).  At the ($10^6$ K, 1.15 $R_{Sun}$) bottom of the corona in our model coronal-hole open flux tube, B = 10 G, $n_e$ = 3 x $10^7$ cm$^{-3}$, and $F_A$ = $10^6$ erg cm$^{-2}$ s$^{-1}$, which values give $\delta B \approx$ 0.3 G and $(\delta B)^2/8\pi \approx$ 3 x $10^{-3}$ dyn cm$^{-2}$ (or erg cm$^{-3}$).  So, at the base of the corona in our coronal-hole model open flux tube, the magnetic pressure of the Alfvenic magnetic-twist waves carrying the needed $10^6$ erg cm$^{-2}$ s$^{-1}$ flux of free magnetic energy is about a third of the thermal pressure of the coronal plasma.  This again suggests that the quasi-hydrostatic quasi-isothermal corona within a few tenths of the bottom of the corona in coronal holes is mostly supported against gravity by the gradient of the thermal plasma pressure, but may be somewhat supported by a gradient in the magnetic wave pressure.

With the above conditions and estimates in hand for the transition region, corona, and solar-wind outflow in our model funnel-ended open flux tube for coronal holes, we now turn to what we expect for how the transition region and corona in the model open flux tube should respond to an increase or decrease in the heating of the flux tube's corona.  We assume the increase or decrease in the coronal heating results from a corresponding increase or decrease in the free-magnetic-energy flux $F_A$ carried into the flux tube's corona by magnetic twist waves. We expect the response of the open flux tube's transition region and corona to be similar to that of the transition region and corona in our conceptual model closed-loop flux tube for quasi-unipolar quiet regions.

If there were a step increase in the energy flux of magnetic twist waves into the open flux tube's corona, the temperature of the corona and hence the conduction heat flux out of the corona into the flux tube's transition region would suddenly increase. Also, there would be a step increase in the magnetic-wave pressure in the flux tube's transition region and corona. We propose that, as in our model closed-loop flux tube for quasi-unipolar quiet regions, initially there would not be enough plasma in the open flux tube's transition region to radiate the increase in incoming heat flux. We suppose that the excess heat flux would again pass down through the transition region into the (T ≈ $10^4$ K) chromosphere to heat and drive "chromospheric-evaporation" upflow, thereby increasing the upflow through the transition region into the flux tube's quasi-hydrostatic lower corona. This would continue until enough new heated-plasma mass were added to the flux tube's corona and transition region that (i) the transition region's radiation would balance the increased heat flux from the corona, (ii) the corona's increased rate of heat gain would again be balanced by the corona's total rate of heat loss, and (iii) the flux tube's transition region and corona would again be in steady-outflow quasi-hydrostatic equilibrium. The flux-tube corona's total rate of heat loss is the sum of its rate of heat loss by its own radiation, its rate of heat loss by conduction to the transition region, and its rate of heat loss to the solar wind. By quasi-hydrostatic equilibrium, we mean that the plasma in the flux tube's slowly outflowing transition region and lower corona is supported against gravity to first order by the sum of the plasma's thermal pressure gradient and the gradient of the magnetic-wave pressure.

Conversely, if there were a step decrease in the energy flux of magnetic twist waves into the open flux tube's corona, the temperature of the corona and hence the conduction heat flux out of the corona into the flux tube's transition region would suddenly decrease. Also, there would be a step decrease in the magnetic wave pressure in the flux tube's transition region and corona. For the new lower temperature and lower wave pressure, there would initially be more plasma in the flux tube's lower corona than could be supported in quasi-hydrostatic equilibrium by the sum of the plasma's thermal pressure gradient and the wave-pressure gradient. There would also initially be more plasma in the flux tube's transition region than could be kept from cooling by the reduced heat flux from the flux tube's corona. Essentially, the reduced heating and reduced wave pressure in the flux-tube's corona would result in cooling excess plasma draining out of the flux tube's corona and transition region down into the chromosphere. This would continue until enough cooled-plasma mass were lost from the flux tube's corona and transition region that (i) the transition region's decreased radiation would balance the decreased heat flux from the corona, (ii) the corona's decreased rate of heat gain would balance the corona's decreased total rate of heat loss, and (iii) the flux tube's transition region and lower corona would again be in steady-outflow quasi-hydrostatic equilibrium.

From the above reasoning, we expect that during an ongoing increase in a corona-hole model open flux tube's coronal heating, there should be increasing upward mass flux though the flux tube's magnetic-funnel-ended transition region into the flux tube's corona. Conversely, during an ongoing decrease in a model open flux tube's coronal heating, there should be decreasing upward mass flux, or possibly downward mass flux, in the flux tube's lower corona and transition region.

In Section 4, we consider possibilities for why EUV plumes are brighter than surrounding magnetic funnels, considering (1) the case of coronal heating being greater in plumes than in surrounding magnetic funnels and (2) the case of coronal heating being less in plumes than in surrounding magnetic funnels.

## 4. The Two Possibilities for the Coronal Heating in EUV Plumes

An EUV plume is born [becomes noticeably bright in coronal [T = (1 – 2) x $10^6$ K] and hotter-transition-region (T ≈ 6 x $10^5$ K) EUV emission] as scattered magnetic flux of both polarities in the lanes and cell interiors of the magnetic network in the quasi-unipolar quiet region or coronal hole is drawn together by supergranular convection flows to make a noticeably robust packed clump of mostly majority-polarity flux at the foot of the growing EUV plume. In a quasi-unipolar quiet region, prior to being packed into the foot flux clump of a plume, most of pre-plume majority-polarity flux is in the funnel feet of non-plume far-reaching closed loops. In a coronal hole, the pre-plume majority-polarity flux is mostly in the funnel feet of non-plume open flux tubes. In either setting, each pre-plume funnel has less magnetic flux than the EUV plume's magnetic funnel will have. The plume funnel is made by the melding of these smaller-magnetic-flux pre-plume magnetic funnels.

There are only the following three general possibilities for what happens to the heating in the coronal extents of the smaller funnels as they meld to become a plume funnel: the coronal heating per unit area of the aggregate magnetic flux in the corona (a) increases, (b) decreases, or (c) does not change. Because the EUV emission from hotter-transition-region-temperature and coronal-temperature plasma in the plume magnetic funnel obviously increases as the plume-foot flux clump coalesces and the plume grows in brightness and width in coronal EUV images, we judge that possibility (c) is highly unlikely, and give it no further consideration. That is, we judge that only the first two possibilities are viable: as the magnetic funnel for the plume is forming, the heating in its coronal extent either (a) increases or (b) decreases. Studies of EUV plumes usually take the conventional view that the first possibility is what occurs in the making of EUV plumes. The present paper takes a broader view. While we do not rule out the first possibility, we propose that the second possibility is plausible and might be true. Due to an observational result that favors the second possibility, we suspect that the coronal heating might be found to decrease in plume births, instead of increasing as is usually thought.

### *4.1. The Commonly Assumed Possibility*

It is commonly assumed that the flow of free magnetic energy that comes from the foot of a forming EUV plume's magnetic funnel and dissipates into heat in the coronal extent of the funnel's flux tube increases as the funnel-foot magnetic flux packs together in the birth of the plume in coronal EUV images. This possibility is suggested by the observation that the brightness (and hence the heating) of the corona is generally greater over areas in which the magnetic field has greater strength and flux in photospheric magnetograms.

From our reasoning in Section 3 about our conceptual model transition region and corona in the magnetic funnels and their coronal extensions in quiet regions and coronal holes, we expect that if the heating in the coronal extension of an EUV plume's magnetic funnel – in quiet regions and in coronal holes – increases as the feet of the funnel's magnetic field are packed into the funnel-foot flux clump while the plume turns on in coronal EUV images, then there should be increasing net upward mass flux in the plume during that time. On that basis, we propose that definite observational determination that there is increasing net upward mass flux in EUV plumes during their births would be a strong indication that the heating in the coronal extensions of EUV plumes increases during the plume's birth.

The observed blueshift from 10 – 25 km s$^{-1}$ upflow of (1 – 3) x 10$^6$ K plasma in EUV plumes in central-disk coronal holes is perhaps a hopeful sign for coronal heating in coronal-hole and quiet-region EUV plumes being greater than in neighboring non-plume magnetic funnels, but that blueshift is not conclusive evidence for EUV plumes having increasing net upward mass flux during their birth. We assume that if the coronal heating increases during a plume's birth, then during the plume's maximum-brightness phase, it remains at a quasi-steady enhanced strength near its strength at the end of the plume's birth, and then decreases to its pre-plume strength as the plume fades out. Hence, for increased coronal heating in EUV plumes, by the reasoning in Section 3, we expect (i) that during a quiet-region EUV plume's maximal phase there is at most much less upward mass flux than during the plume's birth, and (ii) that during a coronal-hole plume's maximal phase, the net upward mass flux remains roughly constant at about the strength attained at the end of the plume's birth. Accurate enough tracking of the net upward mass flux in quiet-region and coronal-hole plumes through their entire lives (from birth onset to the end of decay) is needed to conclusively show whether (i) there actually is more net upward mass flux in the birth of quiet-region EUV plumes than in their maximal phase, and whether (ii) there actually is more net upward mass flux in the maximal phase of coronal-hole EUV plumes than before their birth ends. By our reasoning, observational certification of these two expectations would be strong evidence that coronal heating in EUV plumes increases during plume birth, as is commonly assumed.

For increased coronal heating in EUV plumes, that plumes are outstandingly bright in coronal EUV images is presumably somehow directly or indirectly a consequence of the greater flux of Alfvenic magnetic waves (and/or a greater flux of quasi-static fine-scale magnetic-twist energy in the case of quiet-region EUV plumes) in the plume funnel than in adjacent non-plume magnetic funnels. By our reasoning in Section 3, an increase in the flux of free magnetic energy that dissipates into heat in the corona in a quiet-region or coronal-hole magnetic-funnel flux tube should result in an increase of the plasma density in the flux tube's transition region and corona. That should contribute to that funnel being brighter than surrounding funnels (that have no increase in their coronal heating) in EUV emission such as seen by AIA's 171 Å channel.

Increased coronal heating in EUV plumes requires that as the magnetic flux for a plume is packed into the plume-foot flux clump, the magnetoconvection acting on the feet of that magnetic field generates an increasingly greater upflow of free magnetic energy that goes into and increasingly heats the plume-funnel's transition region and corona. Perhaps that increase in free magnetic energy upflow in the plume funnel could be accomplished by the generation of Alfvenic magnetic waves by turbulence in the downflows between photospheric granule convection cells as proposed by van Ballegooijen & Cranmer (2008), provided the energy flux of the generated waves increases strongly enough with increasing field strength in the funnel-foot flux clump. Another idea is that the increase in the flux of free magnetic energy is a consequence of increasing cancellation of bits of minority-polarity flux that are embedded in or emerge in the interior of the funnel-foot flux clump and/or are pushed into the clump's edges by the supergranular convection flows that keep the clump tightly packed (Raouafi & Stenborg 2014; Wang et al 2016).

### 4.2. The Other Possibility

The other possibility for the coronal heating in EUV plumes is that the coronal heating *decreases* as the foot flux clump coalesces and the plume turns on. This possibility is suggested by the observational finding of Falconer et al (2003) that in central-disk quiet regions far from any active region the coronal luminosity (and presumably the coronal heating) increases roughly in proportion to the total length of the perimeters (the total coast length) of the flux lanes and clumps of the underlying magnetic network. Because the coast length of the network flux in an EUV plume's funnel-foot magnetic flux is less after it is tightly packed together in the clump than before it is packed into the clump, the Falconer et al (2003) result suggests that the coronal heating per unit of magnetic flux in the plume's magnetic funnel decreases as the foot clump's flux coalesces to make the clump. On the basis of the Falconer et al (2003) observational finding, we judge that, instead of increasing, decreasing coronal heating as a plume is born is a plausible possibility and might be true.

From our reasoning in Section 3, we expect that if the heating in the coronal extension of an EUV plume funnel's magnetic field decreases as the feet of the plume funnel's magnetic field coalesce to become the funnel-foot packed flux clump while the plume turns on in coronal EUV images, then (i) in quiet-region EUV plumes there should be increasing downward net mass flux in the plume during that time, and (ii) in coronal-hole EUV plumes there should be decreasing upward net mass flux – or perhaps even downward net mass flux – in the plume during that time. On that basis, we propose that definite observational determination that (i) in quiet-region EUV plumes there is increasing downward net mass flux during their births and (ii) in coronal-hole EUV plumes there is either decreasing upward net mass flux or downward net mass flux during their births, would be a strong indication that the heating in the coronal extensions of EUV plumes decreases during plume birth.

If the coronal heating in EUV plumes decreases as the plume is born, then much of the bright coronal EUV emission from an EUV plume might be from plasma that, due to the decreased heating in the corona, has cooled in the plume flux tube's corona and is draining down through the plume's magnetic funnel. For decreased coronal heating in EUV plumes, it is reasonable that EUV plumes are observed to be more outstandingly bright in AIA 171 Å images (from emission from solar plasmas at temperatures near $6 \times 10^5$ K) than in either AIA 193 Å images (from plasma near $1.6 \times 10^6$ K) or AIA 211 Å images (from plasma near $2 \times 10^6$ K). Plasma that is at temperatures of $(1 – 2) \times 10^6$ K in the coronal extents of the pre-plume magnetic funnels that coalesce to make the plume funnel possibly cools to temperatures near $6 \times 10^5$ K – and is therefore bright in AIA 171 Å images – as it drains down through the plume's magnetic funnel. This possibility is compatible with the plasma temperature and density measured in coronal-hole EUV plumes from EUV spectra. The plume plasma temperature is about $8 \times 10^5$ K, $(2 – 4) \times 10^5$ K cooler than the coronal hole's ambient non-plume corona (Del Zanna et al 2003, 2008; Wilhelm et al 2011; Poletto 2015). Hence, the plume plasma should emit more strongly in AIA's 171 Å passband than in either AIA's 193 Å passband or AIA's 211 Å passband. The plume plasma electron density is $(1 – 2) \times 10^9$ cm$^{-3}$, about 3 times greater than in the ambient coronal hole (Del Zanna et al 2003; Wilhelm et al 2011; Poletto 2015). The plume's cooler temperature and greater density relative to the non-plume coronal hole result in the plume standing out brightly in AIA 171 Å images, as in Figure 2.

We conjecture that the presence of the draining cooling plasma would increase the visibility of the upflowing EUV jetlets that are seen in plume coronal EUV images and often appear to be rooted at the edges of the plume's funnel-foot flux clump. This effect might be why EUV jetlets were discovered first at edges of plume-foot flux clumps (Raouafi & Stenborg 2014) and only later were found to also occur at edges of non-plume-foot flux clumps of the magnetic network (Panesar et al 2018). We further conjecture that the blueshifts seen in EUV plumes in coronal EUV emission lines are mostly from the jetlets, that these blueshifts could mask the smaller redshift of a more pervasive but slower downflow of the draining cooling plasma, and that the pervasive slow downflow could give the plume an increasing downward net mass flux or a decreasing upward net mass flux in the birth of the plume.

Falconer et al (2003) and Moore et al (2011, 2013, 2015) have proposed that coronal heating in quiet regions and coronal holes is mostly driven from the edges of the magnetic network by fine-scale magnetic explosions and that spicules are made by such network-edge magnetic explosions. That scenario is suggested by the Falconer et al (2003) empirical finding that the coronal heating in quiet regions increases in direct proportion to the length of the coast line of the underlying magnetic network. In that scenario, each spicule is driven by the explosive eruption of a flux rope from the sheared-field core of a granule-size magnetic arcade in which flux cancellation at the arcade's polarity inversion line has built the flux rope and triggered its eruption. That scenario is the same as for similar larger magnetic arcades in which flux cancellation builds and triggers a core-field flux rope to erupt and drive a coronal jet spire, plausibly the counterpart of a spicule (e.g., Sterling et al 2015, 2020a; Sterling & Moore 2016; Panesar et al 2016; McGlasson et al 2019). In each of these eruptions, larger or smaller, reconnection of the erupting flux rope's twisted magnetic field with encountered far-reaching or open field launches an Alfvenic magnetic twist wave into the corona on the reconnected far-reaching or open field.

Sterling & Moore (2020) and Sterling et al (2020b) propose that many of these magnetic twist waves might become the magnetic-field switchbacks detected in coronal-hole solar wind by Parker Solar Probe. Here we extend that proposal. We propose (1) that in quiet regions, these waves dissipate in the coronal reaches of closed magnetic loops to power much of the coronal heating in those loops, and (2) that in coronal holes, as the Moore et al (1991) modelling results (for reflection of Alfven waves in coronal holes) suggest, (i) the power spectrum of these waves peaks at a period of about 5 minutes, (ii) the longer-period waves are reflected and trapped within roughly 2 $R_{Sun}$ from Sun center and dissipate to heat the corona there, and (iii) the shorter-period waves escape into the solar wind beyond 2 $R_{Sun}$ to accelerate the solar wind, some surviving as the switchbacks detected by Parker Solar Probe. This idea for the magnetic-network origin of switchbacks is in accord with the Parker Solar Probe observation that switchbacks occur in clusters, called patches, each having about the heliocentric angular width of a supergranule cell of the magnetic network in coronal holes and quiet regions (Bale et al 2019; Fargette et al 2021).

If it turns out that the heating of the coronal extent of EUV plumes increases (instead of decreasing as we think it plausibly could) as plumes are born, that would still allow network-edge spicules and jetlets to be the origin of switchbacks in the solar wind from coronal holes, but would be an indication that the coronal heating in quiet regions and coronal holes is mostly driven by magnetoconvection in some way other than by preparing and triggering network-edge magnetic explosions that plausibly make spicules and jetlets.

## 5. Summary and Discussion

We assume that in the birth of an EUV coronal plume in a quiet region or coronal hole, the heating in the coronal extension of the plume magnetic funnel either (a) always increases or (b) always decreases. Reasoning from simple static or steady conceptual models of the transition region and corona in the magnetic funnels and their coronal extensions in quasi-unipolar quiet regions and coronal holes, we conclude the following. If the coronal heating increases in the birth of EUV plumes, then there should be increasing upward net mass flux in EUV plumes during plume birth. If the coronal heating decreases in the birth of EUV plumes, then there should be (i) downward net mass flux in quiet-region EUV plumes during plume birth, and (ii) decreasing upward net mass flux or even downward net mass flux in coronal-hole EUV plumes during plume birth. Observation of definite increasing upward net mass flux during EUV plume birth in quiet regions and coronal holes would be strong evidence that coronal heating increases in EUV plume births. Observation of (i) definite increasing downward net mass flux in quiet-region EUV plumes during plume birth and (ii) definite decreasing upward net mass flux or downward net mass flux in coronal-hole EUV plumes during plume birth, would be strong evidence that coronal heating decreases in EUV plume births.

Falconer et al (2003) found from quantitative analysis of coronal EUV filtergrams and photospheric magnetograms that the rate of coronal heating in quiet regions is directly proportional to the coastline length of the underlying magnetic network. An EUV plume is born (becomes bright) as many initially scattered smaller majority-polarity flux clumps coalesce to become a tightly packed larger flux clump that is the plume's foot and that has less coastline length than the aggregate coastline length that the pre-plume scattered small flux clumps had. The Falconer (2003) proportional dependence of quiet-region coronal heating on the magnetic network's coastline length suggests that the reduction of plume-foot coastline length during plume birth similarly results in reduction of the plume's coronal heating. On that basis, we think coronal heating in EUV plumes being less than in surrounding non-plume funnels is a plausible possibility.

The Falconer et al (2003) evidence that coronal heating in quiet regions increases or decreases in proportion to the increase or decrease in the coastline length of the magnetic network is compatible with and encourages the idea that coronal heating in quiet regions and coronal heating and solar-wind generation in coronal holes are powered by magnetic twist waves from fine-scale magnetic explosions at the edges of the magnetic network (Falconer et al 3003; Moore et al 1999, 2011, 2015). Observational evidence that coronal heating in EUV plumes is definitely less than in surrounding non-plume magnetic funnels would be compatible with and hence further encourage that idea.

Sterling et al (2020b) have proposed that magnetic field switchbacks observed by Parker Solar Probe (PSP) in the solar wind from coronal holes may be remnants of magnetic twist waves from network-edge explosive events observed as jetlets and spicules. If that turns out to be true, then switchbacks would be signatures of a subset of the network-edge magnetic explosions that might power coronal-hole coronal heating and the solar wind from coronal holes, thereby powering the entire heliosphere (Moore et al 2011).

Even if, as is commonly assumed, coronal heating in EUV plumes is greater than in surrounding non-plume magnetic funnels, it still might be that switchbacks are magnetic twist waves from network-edge magnetic explosions, and network-edge magnetic explosions conceivably still might be the main drivers of coronal heating in quiet regions and coronal heating and solar-wind generation in coronal holes. In any case, increasing coronal heating as an EUV plume is born requires increasing the flux of free magnetic energy into the coronal extent of the plume magnetic funnel. That requires that the free magnetic energy flux generated by magnetoconvection in the plume-foot photospheric magnetic flux increases as the flux coalesces into the plume foot's tightly packed robust flux clump. That conceivably could happen in any one or any combination of the following three ways.

First, perhaps the occurrence rate of magnetic-twist-wave-producing network-edge magnetic explosions increases as the plume-foot magnetic flux coalesces, instead of decreasing as we suspect from Falconer et al (2003). That conceivably could result from enough increase in the rate at which fine-scale minority-polarity flux is swept into the edges of the plume-foot flux clump as it coalesces. In that case, network-edge magnetic explosions might be (i) the main drivers of coronal heating in quiet regions and the main drivers of coronal heating and solar-wind generation in coronal holes, and (ii) the magnetic-network sources of switchbacks.

Second, perhaps many tiny magnetic bipoles no larger than a granule ($< \approx 10^3$ km) continually emerge within the plume-foot flux clump, and explosively generate magnetic twist waves from the clump's interior in addition to the twist waves generated at the clump's edges. In that case, magnetic twist waves from fine-scale magnetic explosions throughout the interiors of network flux patches as well as at their edges might be (i) the main drivers of coronal heating in quiet regions and the main drivers of coronal heating and solar-wind generation in coronal holes, and (ii) the magnetic network sources of switchbacks. The presence or absence of such tiny magnetic bipoles inside network magnetic flux clumps might be established by high-sensitivity ultra-high-resolution magnetograms such as from the Big Bear Solar Observatory Goode Solar Telescope (GST) or the National Solar Observatory Daniel K. Inouye Solar Telescope (DKIST).

Third, the increase in free magnetic energy flux might be generated by magnetoconvection in the interior of the plume-foot robust flux clump via generation of magnetic twist waves by the turbulent downflow between granule convection cells, a la van Ballegooijen & Cranmer (2008). In that case, network-edge magnetic explosions (i) would not be the main magnetic-network power source for coronal heating in quiet regions nor for coronal heating and solar-wind generation in coronal holes, but (ii) could still be the source of some switchbacks.

EUV plumes on the central disk usually stand in an area of enhanced quasi-unipolar magnetic network in a quiet region or coronal hole. That tendency can be seen by examination of the magnetic network within a radius of two supergranules from the plume's flux-clump base for the eight EUV plumes presented by Avallone et al (2018). That tendency plausibly holds for EUV plumes and their surrounding quasi-unipolar network in and around polar coronal holes. If it does, then, since it is plausible that EUV plumes have less coronal heating and less coronal density in the coronal extensions of their magnetic funnels than do adjacent non-EUV-plume magnetic funnels, it is plausible that sometimes a polar EUV plume that is on the limb during a solar eclipse will be seen to stand in **a** dim gap between the feet of two adjacent bright white-light plumes. If there **is** stronger coronal heating in EUV plumes than in adjacent non-plume funnels, then every polar EUV plume that stands at the limb during a solar eclipse should be seen to stand in the foot of a white-light plume.

RLM learned from Peter Sturrock to reduce complex solar astrophysical phenomena to their core physics. RLM and ACS were supported by the NASA Science Mission Directorate's Heliophysics Division by research grants from the Heliophysics Supporting Research program, the Heliophysics Guest Investigators program, and the Heliophysics System Observatory Connect program. NKP acknowledges support from NASA's SDO/AIA contract and NASA's HGI and HSR grants. SKT gratefully acknowledges support by NASA HGI (80NSSC21K0520) and HSR (80NSSC23K0093) grants, and NASA contract NNM07AA01C (Hinode). The referee led us to make the paper clearer.

Figure Captions

Figure 1. Schematic vertical cross-section for our conceptual magnetic-field setup for an EUV plume (a) in a quasi-unipolar quiet region or (b) in a coronal hole.  Both drawings are to scale, each spanning 130,000 km horizontally and vertically above the photosphere.  The setup is depicted for when the polarity of the majority of the photospheric magnetic flux is positive.  For simplicity, in this paper we define the transition region in this magnetic setup to be the plasma at temperatures from $10^4$ K to $10^6$ K in the magnetic funnels.  Panel (a): Each minus sign marks a negative-polarity flux clump in a network cell interior between two contiguous funnel-ended far-reaching coronal loops.  Each plus sign marks the positive-polarity network flux clump at the foot of a magnetic funnel.  Each magnetic funnel and its coronal-loop extension are outlined by a pair of blue curves.  The middle funnel is wider and is rooted in a clump of more and stronger positive flux than are the surrounding funnels that hem in the middle funnel.  The middle funnel is the magnetic funnel of an EUV plume during the plume's maximal luminosity in EUV plasma at hotter-transition-region and coronal temperatures around $10^6$ K.  The other magnetic funnels are non-plume funnels.  Horizontal red dashed lines mark the (T = $10^4$ K) top of the chromosphere between feet of magnetic funnels.  Short red arcs are chromospheric magnetic loops that connect network-cell-interior negative flux clumps to the closest edge of a funnel-foot flux clump.  The horizontal green line low in each funnel marks the (T = $10^4$ K) top of the chromosphere in the funnel.  In the non-plume funnels, the horizontal purple line marks the (T = $10^6$ K) top of the transition region and bottom of the corona.  In the plume funnel, the purple line marks the height that the $10^6$ K level had in that magnetic field before it coalesced to form the plume funnel.  When an EUV plume is discernible in AIA coronal EUV images, its plasma at hotter-transition-region temperatures is striated and extends to heights of twice the plume's width, as in Figure 2.  TR stands for Transition Region.  Panel (b): The magnetic setup and horizontal lines are the same as in Panel (a), except for two differences.  One difference is that instead of being far-reaching closed coronal loops, the coronal extensions of the magnetic funnels are open; they extend out into the solar wind from the coronal hole.  The other difference is that here the $10^6$ K top of the transition region and bottom of the corona is ten times higher than in Panel (a) for quasi-unipolar quiet regions.  In the same way as in Panel (a), the purple line in the plume funnel marks the $10^6$ K height in the plume funnel's magnetic field before it coalesced to form the plume funnel.  As in Figure 2, during its maximal luminosity in emission from plasma at hotter-transition-region temperatures a coronal-hole plume is striated and extends to heights of order the $10^5$ km height of the $10^6$ K level in this schematic.

Figure 2. The cradle-to-grave evolution of a typical EUV coronal plume (from Figure 6 of Avallone et al 2018).  Panel (a): A swath of an AIA 211 Å full-Sun image, centered on the coronal hole of the plume, at the time of the plume's maximum brightness in AIA 171 Å images, at 06:00 UT on 2011 July 6.  The white box outlines the area of the 171 Å images in panel (b).  Panel (b), top row: AIA 171 Å images centered on the plume's foot (1) at the start of the plume's birth, (2) during the plume's growth, (3) during the plume's maximum luminosity in AIA 171 Å images, (4) during the plumes decay, and (5) at the end of the plume's decay. Panel (b), bottom row: HMI magnetograms of the same area at the times of the five 171 Å images, showing that the magnetic flux at the plume's foot coalesces into a compact clump as the plume brightens in 171 Å emission and then disperses as the plume dies out.

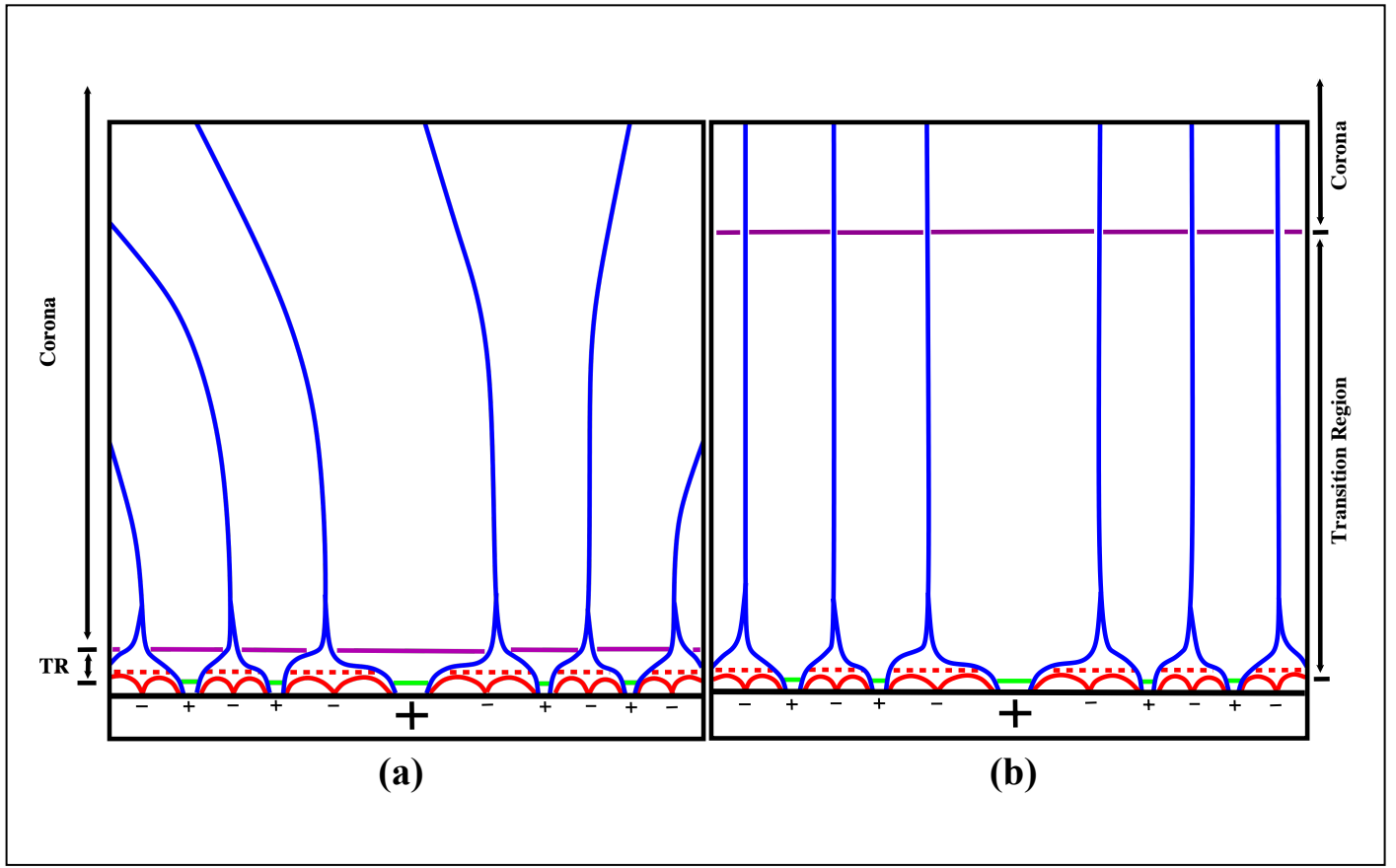

Figure 1

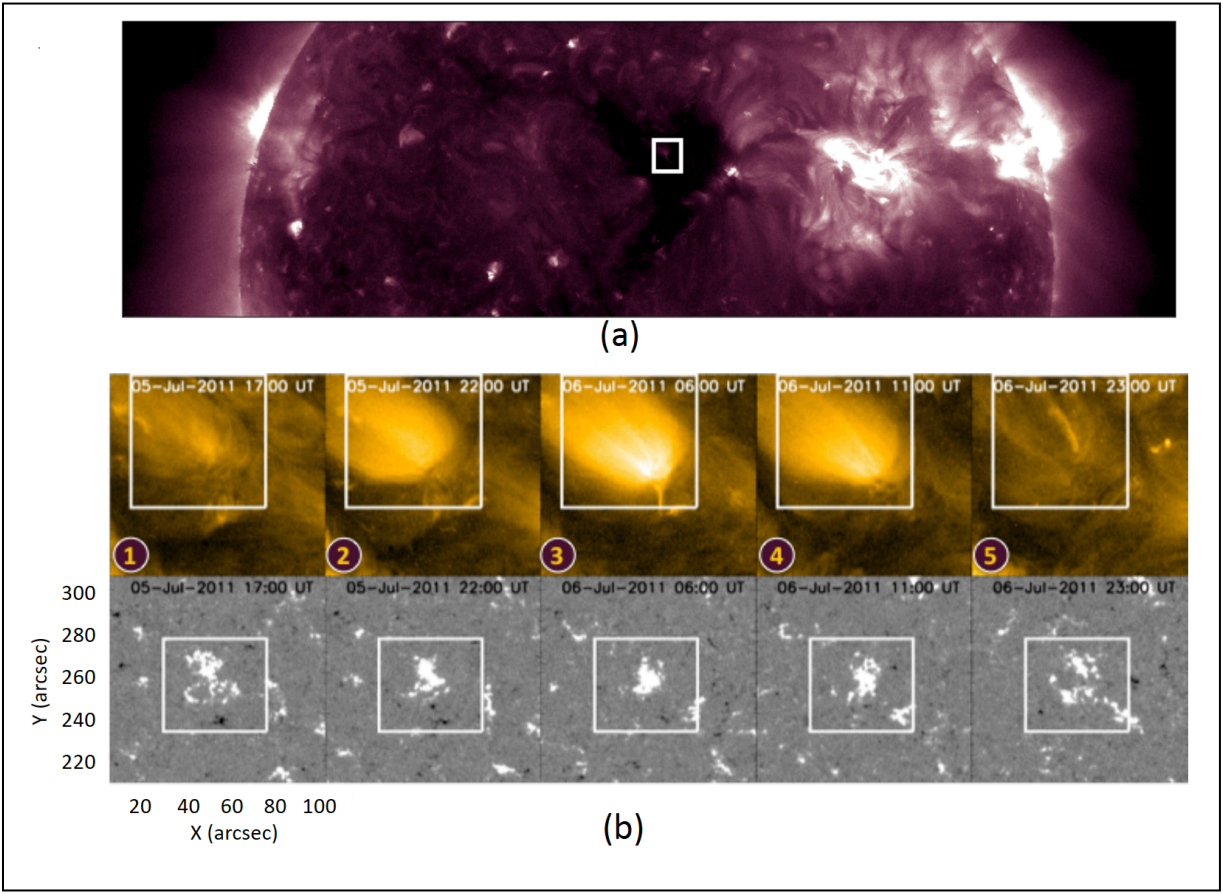

Figure 2